\documentstyle [12pt] {article}
\begin{document}

%%%%Miscellaneous definitions%%%%%%%%%%%%%%%%%%%%%%%%%%%%%%%%%%%%%%%%%%%
\def\asMQ{\alpha_s(M_Q)}
\def\asMc{\alpha_s(M_c)}
\def\asMQpi{{\alpha_s(M_Q) \over \pi}}
\def\chizero{\chi_{c0}}
\def\chione{\chi_{c1}}
\def\chitwo{\chi_{c2}}
\def\chibzero{\chi_{b0}}
\def\chibone{\chi_{b1}}
\def\chibtwo{\chi_{b2}}
\def\Gamhat{{\widehat \Gamma}}
\def\RS{R_{nS}(0)}
\def\RP{R_{nP}'(0)}
\def\qbar{{\bar q}}
\def\Qbar{{\bar Q}}
\def\singlet{{\underline 1}}
\def\octet{{\underline 8}}
\def\QQb{{Q\overline{Q}}}
\def\QQbg{{\QQb g}}
\def\ket#1{{|#1\rangle}}
\def\O{{\cal O}}
\def\S{{\rm S}}
\def\P{{\rm P}}

%%%%Start of Text%%%%%%%%%%%%%%%%%%%%%%%%%%%%%%%%%%%%%%%%%%%%%%%%%%%%%%%%%%%%
\pagestyle{empty}
\rightline{\vbox{
\halign{&#\hfil\cr
&ANL-HEP-PR-92-30\cr
&NUHEP-TH-92-4\cr
&April 1992\cr}}}
\bigskip
\bigskip
\bigskip
{\Large\bf
\centerline{Rigorous QCD Predictions}
\centerline{for Decays of P-Wave Quarkonia}}
\bigskip
\normalsize
\centerline{Geoffrey T. Bodwin}
\centerline{\sl High Energy Physics Division, Argonne National Laboratory,
    Argonne, IL 60439}
\bigskip

\centerline{Eric Braaten}
\centerline{\sl Department of Physics and Astronomy, Northwestern University,
    Evanston, IL 60208}
\bigskip

\centerline{G. Peter Lepage}
\centerline{\sl Newman Laboratory of Nuclear Studies, Cornell University,
    Ithaca, NY 14853}
\bigskip

\begin{abstract}
Rigorous QCD predictions for decay rates of the P-wave states of heavy
quarkonia are presented.  They are based on a new factorization theorem
which is valid to leading order in the heavy quark velocity and to all
orders in the running coupling constant of QCD.  The decay rates for all
four P states into light hadronic or electromagnetic final states are
expressed in terms of two phenomenological parameters, whose
coefficients are perturbatively calculable.  Logarithms of the binding
energy encountered in previous perturbative calculations of P-wave
decays are factored into a phenomenological parameter that is related to
the probability for the heavy quark-antiquark pair to be in a
color-octet S-wave state. Applying these predictions to charmonium, we
use measured decay rates for the $\chione$ and $\chitwo$ to predict the
decay rates of the $\chizero$ and $h_c$.
\end{abstract}
\vfill\eject\pagestyle{plain}\setcounter{page}{1}

One of the earliest applications of perturbative QCD was the calculation
of the light hadronic and electromagnetic decay rates of S-wave
quarkonia \cite{aprg}.
That calculation was based on the assumption that the annihilation
of the heavy quark and antiquark is a short distance process which,
because of the asymptotic freedom of QCD, can be
computed in perturbation theory. It was assumed that
nonperturbative effects could
be factored into $\RS$, the nonrelativistic wavefunction
at the origin.  This assumption has been supported by
subsequent calculations beyond leading order \cite{barba,mlp}.
Taking a similar approach in the case of
P-wave states \cite{barbb},
one might expect to be able to calculate their
decay rates into light hadrons
in terms of a single nonperturbative input $\RP$, the derivative of the
nonrelativistic wavefunction at the origin.
Unfortunately explicit calculations at order
$\alpha_s^3$ (leading order for $^3\P_1$ and $^1\P_1$ (Ref.~\cite{barbc})
and next-to-leading order for $^3\P_0$ and $^3\P_2$ (Ref.~\cite{barbd}))
reveal infrared divergences ---
a clear indication of sensitivity to nonperturbative effects
beyond those contained in $\RP$.
In previous phenomenological applications, the divergence has been replaced
by a logarithm of the binding energy or
the confinement radius \cite{barbd}
or the radius of the bound state \cite{kmrr}.
None of these prescriptions has any fundamental justification.

A rigorous treatment of the P-wave decays requires a clean separation
between short distance effects, which can be calculated as a
perturbation series in the running coupling constant of QCD,
and long distance effects.  The long distance effects
must either be calculated with some nonperturbative method, such as
lattice QCD, or else absorbed into a small set of parameters
that can be determined phenomenologically.
In this paper, we present predictions for P-wave decays based on a
rigorous QCD analysis.
The QCD predictions are based on factorization theorems \cite{bbl}
that will only be justified at an intuitive level in this paper, since
our focus will be on their phenomenological implications.
The decay rates of the four P states into light hadrons, as well as the
decay rates of the $^3\P_0$ and $^3\P_2$ states into two photons,
are all predicted in terms of two phenomenological parameters.
In addition to $\RP$, there is a second parameter associated
with the probability for the heavy
quark and antiquark to be in a color-octet S state.
When these predictions are combined with
equally rigorous relations between the rates
for radiative transitions between P and S states,
their predictive power is quite remarkable.
{}From the decay rates of the $^3\P_2$ and $^3\P_1$
states of the first radial excitation of quarkonium,
we are able to predict the inclusive decay rates
of the $^3\P_0$ and $^1\P_1$ states.
For charmonium, some of our predictions differ significantly
from the presently accepted results.

The $n$'th radial excitation of heavy quarkonium is split
into angular-momentum states $n(^{2S+1}L_J)$
with parity $(-1)^{L+1}$ and charge conjugation $(-1)^{L+S}$.
Our basic results are factorization formulae
for the decay rates of these states into light hadronic or
electromagnetic final states.  The factorization formulae are valid
to leading order in $v^2$, where $v$ is the typical velocity
of the heavy quark,  and to all orders in the QCD coupling constant
$\asMQ$.  For S-wave $(L = 0)$ and P-wave $(L = 1)$ states,
they have the schematic form
\begin{equation}
	{
\Gamma \left( n(^{2S+1}\S) \rightarrow X \right) \; = \;
G_1(n) \; \Gamhat_1
\left( Q \Qbar (^{2S+1}\S) \rightarrow X \right) \; ,
}
\label{factS}
\end{equation}
\begin{equation}{
\Gamma \left( n(^{2S+1}\P) \rightarrow X \right) \; = \;
H_1(n) \; \Gamhat_1
\left( Q \Qbar (^{2S+1}\P) \rightarrow X \right) \;
+ \;  H_8(n) \; \Gamhat_8
\left( Q \Qbar (^{2S+1}\S) \rightarrow X \right) \; .
}
\label{factP}
\end{equation}
In (\ref{factS}) and (\ref{factP}),
all nonperturbative effects are factored into
the parameters $G_1(n)$, $H_1(n)$, and $H_8(n)$.  They
depend on the radial quantum number $n$, but
are independent of the total spin $S$ and total
angular momentum $J$ to leading order in $v^2$.
The spin-independence follows from the fact that the
QCD interactions of heavy quarks are independent
of the spin of the quark, up to relativistic corrections.
The factors $\Gamhat_1$ and $\Gamhat_8$
are hard subprocess rates for the annihilation of the heavy quark and
antiquark at threshold, with the $Q \Qbar$ in a $^{2S+1}L$
angular-momentum state and in a color-singlet state for $\Gamhat_1$
and a color-octet state for $\Gamhat_8$.
The subprocess rates are calculable as a perturbation series in $\asMQ$,
the QCD running coupling constant evaluated at the heavy-quark mass.
The parameters $G_1$, $H_1$, and $H_8$
are proportional to the probabilities for the
bound state to contain a $Q \Qbar$ pair in a color-singlet S-wave,
a color-singlet P-wave,
and a color-octet S-wave state, respectively,
with separation $r \rightarrow 0$.
They may simply be taken as
phenomenological parameters to be determined by experiment, but
they can also be given
rigorous nonperturbative definitions \cite{bbl}
and can therefore be measured using lattice simulations of QCD.
In the factorization formula (\ref{factP}),
$H_8$ and $\Gamhat_1$ depend on an arbitrary
factorization scale $\mu$ in such a way that the complete decay rate is
independent of $\mu$.  We have implicitly set $\mu = M_Q$.

The color-octet contributions to P-wave decays ($H_8\Gamhat_8$)
may seem peculiar if one is used to thinking of mesons as color-singlet
$\QQb$ states.  However, any meson is a superposition of many
components, involving any number of quarks and gluons:
\begin{equation}
{
 \ket M \; = \; \psi_\QQb \ket\QQb
	\; + \;  \psi_\QQbg \ket\QQbg \; + \; \cdots .
}
\label{ketM}
\end{equation}
In some of these components, notably $\ket \QQbg$,
the heavy quark and antiquark are in a color-octet state.
The color-singlet and color-octet pieces of the our
factored decay rates
represent contributions coming from $\QQb$ and $\QQbg$
components of the meson respectively.
The probability carried by the $\QQbg$
state (and higher states) is of order $v^2$ --- heavy quarks don't
easily radiate a gluon. Consequently such components can be neglected
for many applications, including S-wave annihilations.
However, the $\ket\QQb$ contribution to annihilations of P-wave quarkonium
is suppressed by $v^2$, owing to the angular-momentum barrier, which pushes the
quarks apart.  Furthermore, in the $\QQbg$ component of P-wave quarkonium,
the quark and antiquark can be in an S-wave state, with
no angular-momentum barrier to hinder their annihilation.
The $\QQbg$ contribution to the annihilation therefore competes
with, and in some cases even dominates, that coming from $\QQb$.
The enhanced role of the $\QQbg$ component
makes the decays of P states particularly interesting:
they are among the very
few processes in quarkonium physics that give us a glimpse of
physics beyond the simple quark potential model.

To leading order in $v^2$, $H_1$ is directly related to the
nonrelativistic wavefunction $\psi_\QQb = R_{nP}(r) Y_{1m}(\hat r)$:
\begin{equation}
{
H_1(n) \; \approx \; {9\over 2\pi} {|R^\prime_{nP}(0)|^2 \over M_Q^4} .
}
\label{hsinglet}
\end{equation}
There is no simple formula for $H_8$ in terms of $R_{nP}$ since $H_8$
is determined by the $\QQbg$ wavefunction~$\psi_\QQbg$ rather than by
$\psi_\QQb$. However, in the limit of very large quark mass, $H_8$ is
dominated by $\QQbg$ configurations in which the gluon has large
momentum, and an approximate perturbative
relation \cite{bbl} can be obtained between $H_1$ and $H_8$:
\begin{equation}
{
 H_8(n) \; \approx \; {16 \over 27 \beta_0}
\ln  \left( {\alpha_s(E_n) \over \alpha_s(M_Q)} \right) \, H_1(n) \; ,
}
\label{hoctet}
\end{equation}
where $E_n$ is the binding energy, $\beta_0 = (33-2n_f)/6$,
and $n_f$ is the number of light flavors.
The constant accompanying the logarithm in (\ref{hoctet})
cannot be computed perturbatively and is
probably quite important for charmonium and bottomonium.

If in calculating the gluon emission process that gives (\ref{hoctet}),
one neglects the running of the coupling constant,
then the perturbative expression for $H_8$ reduces to
\begin{equation}
{
 H_8(n) \; \sim \; {16 \over 27 \pi} \alpha_s
\ln \left( M_Q \over E_n \right) \, H_1(n) \; .
}
\label{hoctetb}
\end{equation}
This logarithm of the binding energy is precisely the infrared
divergence that was found in previous
(nonrigorous) perturbative analyses of P-wave decays \cite{barbc,barbd}.
Infrared divergences arose in this earlier work
because the $\QQbg$ component of the meson was neglected.  In our
analysis the infrared sensitivity is factored into the nonperturbative
parameter $H_8$, so the subprocess rate
$\Gamhat_1$ involves only hard contributions.

For the decay rates of the P states into light hadrons,
the factorization formula (\ref{factP}) in more explicit form is
\begin{eqnarray}
\Gamma \left( n(^3\P_J) \rightarrow {\it l. h.} \right) &=&
H_1(n) \; \Gamhat_1
\left( Q \Qbar (^3\P_J) \rightarrow {\it partons} \right) \nonumber\\
&&\mbox{}+ \; H_8(n) \; \Gamhat_8
\left( Q \Qbar (^3\S_1) \rightarrow {\it partons} \right) \; ,
\; J\; = \; 0, 1, 2 \; ,
\label{facttrip}
\end{eqnarray}
\begin{eqnarray}
\Gamma \left( n(^1\P_1) \rightarrow {\it l. h.} \right) &=&
H_1(n) \; \Gamhat_1
\left( Q \Qbar (^1\P_1) \rightarrow {\it partons} \right) \nonumber\\
&&\mbox{}+ \;  H_8(n) \; \Gamhat_8
\left( Q \Qbar (^1\S_0) \rightarrow {\it partons} \right) \; ,
\label{factsing}
\end{eqnarray}
where $``{\it l. h.}"$ on the left side of (\ref{facttrip}) or
(\ref{factsing})
represents all final states consisting of light hadrons
and $``{\it partons}"$ on the right side
represents perturbative final states consisting of gluons and light
quark-antiquark pairs.  Note that, for the $^3\P_J$ decays,
the second term on the right side of (\ref{facttrip}) is independent of $J$.
For the decay rates of the $^3\P_0$ and $^3\P_2$ states
into two photons, the factorization formula (\ref{facttrip}) simplifies
to the color-singlet term only,
because color conservation forbids the annihilation
of the $Q \Qbar$ in a color-octet state:
\begin{equation}
{
\Gamma \left( n(^3\P_J) \rightarrow \gamma \gamma \right) \; = \;
H_1(n) \; \Gamhat_1
\left( Q \Qbar (^3\P_J) \rightarrow \gamma \gamma \right) \; ,
\; J \; = \; 0, 2 \; .
}
\label{factem}
\end{equation}

When applied to decays into a hard photon plus light hadrons,
the factorization formula (\ref{factsing}) has remarkable
implications.  The color-octet
term allows the $^1\P_1$ state to decay into a hard photon
plus light hadrons at order $\alpha \asMQ$ through
the subprocess $Q \Qbar (^1\S_0) \rightarrow  \gamma g$.
This decay produces a final state consisting of
a hard photon recoiling against a hard gluon jet
and against the hadrons from the fragmentation of the gluon
in the $Q \Qbar g$ state.
It is possible that this dramatic decay mode of the $^1\P_1$ state
could serve as an experimental signature for this particle.

The leading subprocess rates $\Gamhat_1$ and $\Gamhat_8$
for light-hadronic and electromagnetic decays of P states
are listed in Table 1.
They can be extracted from previous calculations
of P-wave decays \cite{barbb,barbc,barbd} by
using the expressions (\ref{hsinglet})
and (\ref{hoctetb}) for $H_1$ and $H_8$.
Since next-to-leading-order corrections in $\alpha_s$ have not
been computed for the $\Gamhat_8$'s, we work only to leading order
throughout. However, we note that, since $H_1$ and
$H_8$ are independent parameters, a higher-order contribution involving
$H_1$ could, in principle, be numerically important compared to a
leading-order contribution involving $H_8$.

The two nonperturbative parameters $H_1$ and $H_8$ can be
obtained by measuring the decay rates into light hadrons
of the $^3\P_1$ and $^3\P_2$ states,
which are the two P states most accessible to experiment.
At leading order in $v^2$ and $\asMQ$, they are
\begin{equation}
{
H_1 \; \simeq \;
{45 \over 16 \pi}
{\Gamma \left( ^3\P_2 \rightarrow {\it l. h.} \right)
- \Gamma \left( ^3\P_1 \rightarrow {\it l. h.} \right)
\over \alpha_s^2(M_Q)} \; ,
}
\label{hone}
\end{equation}
\begin{equation}
{
H_8 \; \simeq \;
{3 \over \pi n_f}
{\Gamma \left( ^3\P_1 \rightarrow {\it l. h.} \right)
\over \alpha_s^2(M_Q)} \; .
}
\label{height}
\end{equation}

For applications at leading order in $\asMQ$, it is convenient
to eliminate $H_1$ and $H_8$ to obtain direct relations
between decay rates.
{}From (\ref{facttrip}) and (\ref{factem}), we get the following relations
between the decay rates of the spin-triplet states, valid to leading order
in $v^2$ and in $\asMQ$:
\begin{equation}{
{ \Gamma( ^3\P_0 \rightarrow {\it l. h.} )
	\; - \; \Gamma( ^3\P_1 \rightarrow {\it l. h.} )
\over \Gamma( ^3\P_2 \rightarrow {\it l. h.} )
	\; - \; \Gamma( ^3\P_1 \rightarrow {\it l. h.} ) }
\; \simeq \; {15 \over 4}  \; ,
}
\label{ratiob}
\end{equation}
\begin{equation}
{
{ \Gamma( ^3\P_0 \rightarrow \gamma \gamma )
\over \Gamma( ^3\P_2 \rightarrow {\it l. h.} )
	\; - \; \Gamma( ^3\P_1 \rightarrow {\it l. h.} ) }
\; \simeq \; {135 \over 8} e_Q^4 \left( {\alpha \over \asMQ} \right)^2 \; ,
}
\label{ratioc}
\end{equation}
\begin{equation}
{
{ \Gamma( ^3\P_2 \rightarrow \gamma \gamma )
\over \Gamma( ^3\P_2 \rightarrow {\it l. h.} )
	\; - \; \Gamma( ^3\P_1 \rightarrow {\it l. h.} ) }
\; \simeq \; {9 \over 2} e_Q^4 \left( {\alpha \over \asMQ} \right)^2 \; .
}
\label{ratiod}
\end{equation}
These differ from previous predictions \cite{barbb},
which can be obtained by setting
$\Gamma( ^3\P_1 \rightarrow {\it l. h.} ) = 0$.
At leading order, the decays of the two spin-1 states involve only the
color-octet term, so the ratios of the decay rates are
simply the ratios of the subprocess rates $\Gamhat_8$ in Table 1:
\begin{equation}
{
{ \Gamma( ^1\P_1 \rightarrow {\it l. h.} )
	\over \Gamma( ^3\P_1 \rightarrow {\it l. h.} ) }
\; \simeq \; {5 \over 2 n_f}   \; ,
}
\label{ratioa}
\end{equation}
\begin{equation}
{
{ \Gamma( ^1\P_1 \rightarrow \gamma + {\it l. h.} )
	\over \Gamma( ^3\P_1 \rightarrow {\it l. h.} ) }
\; \simeq \; {6 \over n_f} e_Q^2 {\alpha \over \asMQ}  \; .
}
\label{ratioe}
\end{equation}
The prediction (\ref{ratioa}) was first made by Barbieri {\it et
al.} \cite{barbc}.

There are also relations among the radiative transitions
between P and S states that
are accurate up to relativistic corrections of order $v^2$
(Ref.~\cite{mb}).
The predictions for the decay rates of P states into S states plus a photon are
\begin{equation}
{
{\Gamma( ^1\P_1 \rightarrow \gamma \; ^1\S_0 ) \over E_\gamma^3}
\; \simeq \;
{\Gamma( ^3\P_J \rightarrow \gamma \; ^3\S_1 ) \over E_\gamma^3},
\; J = 0, 1, 2 \; ,
}
\label{eqptos}
\end{equation}
where $E_\gamma$ in the denominator is understood to be the energy
of the photon for the transition in the numerator.  In terms of
the masses $M_P$ and $M_S$ of the bound states,
$E_\gamma = (M_P^2 - M_S^2)/(2 M_P)$.

We now apply our QCD predictions to the charmonium system.
For the first radial excitation,
the $ ^3\P_J$ states are called $\chi_{cJ}$ and the $ ^1\P_1$ state is
called the $h_c$.  We use measured decay rates of the
$\chione$ and $\chitwo$ to predict the
inclusive decay rates of the $\chizero$ and $h_c$.
It is important to have reasonable estimates for the theoretical errors
in our predictions if they are to be compared with experimental results.
The two main sources of theoretical error are relativistic corrections
and higher-order perturbative corrections.
In potential models of charmonium,
the average value of $v^2$ is found to be about 0.23 (Ref.~\cite{bt}).
Since our factorization formulae
are valid only to leading order in $v^2$,
we expect an error on the order of 20\%
due to relativistic effects.  Similarly, we
expect deviations on the order of 20\% from the equalities
(\ref{eqptos})
involving the radiative transition rates.
We can estimate the perturbative error from
the size of the perturbative corrections in other bound state calculations.
Based on a number of next-to-leading-order calculations for S-wave bound
states \cite{kmrr}, we estimate the perturbative error
to be  $4 \asMc/\pi$, where $M_c$ is the
mass of the charm quark.  To avoid the ambiguity in the value of
$M_c$, we determine $\asMc$ by
taking the coupling constant $\alpha_s(M_b) = 0.179 \pm 0.009$ extracted
from bottomonium decays \cite{kmrr}
and evolving it down to the scale $M_c$.  This does not require
that we know
$M_c$ and $M_b$ separately, but only their ratio,
for which we use the ratio of the $J/\psi$ and $\Upsilon$ masses:
$M_c/M_b \simeq 0.33$. The resulting value of the
coupling constant is $\asMc = 0.25 \pm 0.02$.
Our estimate of the perturbative error is therefore $30\%$.
We treat the $8\%$ error in
the value of $\asMc$ itself as an experimental error.
We take the QED coupling at the scale $M_c$ to be $\alpha = 1/133.3$.

We assume in our analysis that
the decays into light hadrons and the radiative transitions
to $J/\psi$ or $\eta_c$ are the only decay modes  which contribute
appreciably to the total decay rates of $\chi_{cJ}$ and $h_c$.
In particular, we neglect pionic transitions of the P states
to the S states, of which the most important decay modes
should be $J/\psi + \pi \pi$ and $\eta_c + \pi \pi$.
The rate for the particular decay $h_c \rightarrow J/\psi + \pi \pi$
has been estimated within a well-developed phenomenological
framework \cite{kty}
to be on the order of 6 keV, and the other two-pion
transition rates should be of
the same order of magnitude.  The errors due to neglecting these
contributions to the total
decay rates are negligible compared to other errors.

Precision measurements of the total decay rates of the
$^3\P_1$ state $\chione$ and the $^3\P_2$ state $\chitwo$ have
recently been carried out at Fermilab by
the E760 collaboration \cite{rosen}. Their results,
with statistical and systematic errors added in quadrature, are
listed as input values in the first column of Table 2.
Previous experiments have measured
the branching fractions for the radiative transitions
of the $\chione$ and $\chitwo$ into the
$J/\psi$ \cite{pdg}, and they
are listed as input values in the first column of Table 3.
We use the radiative branching fractions
and the total decay rates to obtain the partial rates given in Table 2
for light-hadronic and radiative decays of the $\chione$ and $\chitwo$.

Inserting the partial rates into light hadrons into (\ref{hone}) and
(\ref{height}),
we determine the nonperturbative parameters for P-wave decays
of charmonium to be
\begin{equation}
{
H_1 \; \simeq \; 15.3 \pm 3.7 \; {\rm MeV} \; (\pm 36 \%) \;,
}
\label{Honeexp}
\end{equation}
\begin{equation}
{
H_8 \; \simeq \; 3.26 \pm 0.73 \; {\rm MeV} \; (\pm 36 \% ) \;.
}
\label{Heightexp}
\end{equation}
Here, and throughout the remainder of this paper,
the first error is from the uncertainties in
our experimental inputs and the second error
is our estimate of the theoretical uncertainty.
The theoretical error is computed by combining
the relativistic error of $20\%$ and the perturbative error of $30\%$,
using the standard formulae for propagating independent errors.
The ratio of the two nonperturbative parameters
in (\ref{Honeexp}) and (\ref{Heightexp}) is $H_8/H_1 \simeq 0.21$.
This is roughly consistent with the value that one would obtain from the
leading-log perturbative expression (\ref{hoctet})
by arbitrarily setting $\alpha_s(E_n) \sim 1$.  However, there
is no {\it a priori} justification for using the leading-log approximation
at such small values of the heavy quark-mass.

We proceed to calculate the total decay rates of the
$\chizero$ and $h_c$.
The experimental values for the radiative transition rates of
$\chione$ and $\chitwo$ in Table 2 are
in good agreement with the theoretical prediction (\ref{eqptos}).
This gives us confidence in our predictions in Table 2 for the
radiative transition rates of $\chizero$ and $h_c$.
In calculating the radiative transition rate of the $h_c$,
we have assumed that its mass is at the center of gravity of the
nine spin states of $\chizero$, $\chione$, and $\chitwo$:
$M_{h_c} = 3525$ MeV.
The predictions in Table 2 for the decay rates of $\chizero$ and $h_c$
into light hadrons are obtained  by using (\ref{ratiob}) and
(\ref{ratioa}).
Adding the light-hadronic and radiative decay rates,
we obtain the total decay rates given in Table 2.
The prediction for the $\chizero$ differs
from the presently accepted value \cite{pdg}
of $(14 \pm 5)$ MeV by several standard deviations.
Our calculations suggest a real discrepancy
between this measurement of the $\chizero$ decay rate
and the experimental data on the
$\chione$ and $\chitwo$ that we have used as our input.

In Table 3, we present our predictions for the branching
fractions for decays of the charmonium P states into final
states containing photons.
The branching fractions for $\chizero$ and $\chitwo$ into two photons
are calculated from (\ref{ratioc}) and (\ref{ratiod}).
The branching fraction for the $h_c$ to decay
into a hard photon plus light hadrons is calculated using (\ref{ratioe}).
Our prediction for the radiative branching fraction of the $\chizero$
is significantly larger than
the accepted value \cite{pdg} of $0.0066 \pm 0.0018$.
Our prediction for the branching fraction of the  $\chitwo$
into two photons is considerably smaller than
the accepted value \cite{pdg} of $(11 \pm 6) \times 10^{-4}$.
A rather tantalizing prediction is that the
branching fraction for the $h_c$ to decay into a hard
photon plus light hadrons is about $2\%$.  This is
large enough that it may be possible to detect the
decay of the $h_c$ by observing the resulting hard photon.

P-wave annihilations provide unique information on the dynamical role of
the gluon in determining hadron structure.  They challenge us to go
beyond the quark-potential picture in modelling hadrons.  In this paper
we have outlined the first rigorous formalism for describing these
processes in QCD.  We have applied the formalism in a detailed analysis
of charmonium P-wave decays.  Within our formalism it is possible to
refine systematically the theoretical predictions for the decay rates,
both by computing higher-order perturbative contributions, and also by
using lattice simulations to compute the nonperturbative parameters. The
promise of significant improvements in both theory and experiment make
the P-wave decays important testing grounds for ideas about both
perturbative and nonperturbative QCD.

This work was supported in part by the U.S. Department of Energy,
Division of High Energy Physics,
under Contract W-31-109-ENG-38 and under Grant DE-FG02-91-ER40684,
and by the National Science Foundation.
\vfill\eject

\vfill\eject

\noindent{\Large\bf Table Captions}
\begin{enumerate}
\item Subprocess rates $\Gamhat_1$ and $\Gamhat_8$ for decays of
P-wave quarkonium states.
\item Predictions for total and partial decay rates of P-wave charmonium
states.
\item Predictions for branching fractions of P-wave charmonium states.
\end{enumerate}
\vfill\eject

\noindent{\Large\bf Tables}
\vskip 1cm
\renewcommand{\arraystretch}{1.5}
%%%%%%%%%%%%%%%%%%%%% TABLE 1 %%%%%%%%%%%%%%%%%%%%%%%%%%%%%%%%%%%%%%%%
\begin{tabular}{|c|c|c|c|c|}
\hline
decay mode&$\Gamhat_1$&\vbox{\hbox{\strut color-singlet}
\hbox{\strut subprocess}} &
	 $\Gamhat_8$ & \vbox{\hbox{\strut color-octet}
\hbox{\strut subprocess}} \\
\hline
 $^3\P_0 \rightarrow {\it l.h.}$ &
 	$(4\pi/3) \alpha_s^2$ & $^3\P_0 \rightarrow gg$ &
 	$(\pi n_f/3) \alpha_s^2$ & $^3\S_1 \rightarrow q \qbar$ \\
 $^3\P_1 \rightarrow {\it l.h.}$ &
  	$\O(\alpha_s^3)$ & &
 	$(\pi n_f/3) \alpha_s^2$ & $^3\S_1 \rightarrow q \qbar$ \\
 $^3\P_2 \rightarrow {\it l.h.}$ &
 	$(16\pi/45) \alpha_s^2$ & $^3\P_2 \rightarrow gg$ &
 	$(\pi n_f/3) \alpha_s^2$ & $^3\S_1 \rightarrow q \qbar$ \\
 $^1\P_1 \rightarrow {\it l.h.}$ &
  	$\O(\alpha_s^3)$ & &
 	$(5 \pi/6) \alpha_s^2$ & $^1\S_0 \rightarrow gg$ \\
\hline
 $^3\P_J \rightarrow \gamma + {\it l.h.}$ &
 	$\O(\alpha \alpha_s^2)$ & &
 	$\O(\alpha \alpha_s^2)$ & \\
 $^1\P_1 \rightarrow \gamma + {\it l.h.}$ &
  	$\O(\alpha \alpha_s^2)$ & &
 	$2 \pi e_Q^2 \alpha \alpha_s$ & $^1\S_0 \rightarrow \gamma g$ \\
\hline
 $^3\P_0 \rightarrow \gamma \gamma$ &
 	$6 \pi e_Q^4 \alpha^2$ & $^3\P_0 \rightarrow \gamma \gamma$ &
 	$0$ & \\
 $^3\P_2 \rightarrow \gamma \gamma$ &
 	$(8\pi/5) e_Q^4 \alpha^2$ & $^3\P_2 \rightarrow \gamma \gamma$ &
	$0$ & \\
\hline
\end{tabular}
 \vskip 0.7 true cm
 \vbox{
	\baselineskip=6 mm
 	\noindent
 	\centerline{Table 1}
 }
 \vskip 1 cm
%%%%%%%%%%%%%%%%%%%%%%%%%%%%%%%%%%%%%%%%%%%%%%%%%%%%%%%%%%%%%%%%%%%%%%%

%%%%%%%%%%%%%%%%%%%%% TABLE 2 %%%%%%%%%%%%%%%%%%%%%%%%%%%%%%%%%%%%%%%%
\begin{tabular}{|c|c|c|c|}
\hline
&$\Gamma$ in MeV&$\Gamma({\it l.h.})$&$\Gamma(\gamma J/\psi),
\Gamma(\gamma \eta_c)$\\
\hline
$\chizero$&$4.8 \pm 0.7 \;(\pm 35\%)$&$4.7 \pm 0.7 \; (\pm 36\%)$&$0.099
\pm 0.010 \; (\pm 20\%)$\\
$\chione$&INPUT: $0.88 \pm 0.14$&$0.64 \pm 0.10 $&$0.240 \pm 0.041$\\
$\chitwo$&INPUT: $1.98 \pm 0.18$&$1.71 \pm 0.16 $&$0.267 \pm 0.033$\\
$ h_c $&$0.98 \pm 0.09  \; (\pm 22\%)$&$0.53 \pm 0.08 \; (\pm 36\%)$&$0.45
\pm 0.05 \; (\pm 20\%)$\\
\hline
\end{tabular}
 \vskip 0.7 true cm
 \vbox{
 \baselineskip=6 mm
 \noindent
 \centerline{Table 2}
 }
 \vskip 1 cm
%%%%%%%%%%%%%%%%%%%%%%%%%%%%%%%%%%%%%%%%%%%%%%%%%%%%%%%%%%%%%%%%%%%%%%%
\vfill\eject

%%%%%%%%%%%%%%%%%%%%% TABLE 3 %%%%%%%%%%%%%%%%%%%%%%%%%%%%%%%%%%%%%%%%
\begin{tabular}{|c|c|c|c|}
\hline
&	 $B(\gamma J/\psi), B(\gamma \eta_c)$ &
	 $B(\gamma + {\it l.h.})$ &
	 $B(\gamma \gamma)$ \\
\hline
$\chizero$ &
 	$0.021 \pm 0.004 \; (\pm 40\%)$ &
	$ $ &
 	$(6.8 \pm 1.9) \times 10^{-4} \; (\pm 50\%)$ \\
 $\chione$ &
 	INPUT: $0.273 \pm 0.016$ &
	$ $ &
	$0$ \\
 $\chitwo$ &
	INPUT: $0.135 \pm 0.011$ &
	 $ $ &
	 $(4.1 \pm 1.1) \times 10^{-4} \; (\pm 36\%)$ \\
 $ h_c $ &
 	$0.46 \pm 0.05 \; (\pm 22\%)$ &
 	$0.017 \pm 0.003 \; (\pm 42\%)$ &
	$0$	\\
\hline
\end{tabular}
 \vskip 0.7 true cm
 \vbox{
 \baselineskip=6 mm
 \noindent
 \centerline{Table 3}
 }
 \vskip 1 cm
%%%%%%%%%%%%%%%%%%%%%%%%%%%%%%%%%%%%%%%%%%%%%%%%%%%%%%%%%%%%%%%%%%%%%%%

\end{document}